\documentclass[twocolumn]{aastex62}
\usepackage{amsmath,amssymb}
\usepackage{graphicx}
\usepackage{bm}
\usepackage{here}

\begin{document}

\title{\Large Coronae of Zero/Low-Metal Low-Mass Stars}
\author{\normalsize Haruka Washinoue} 
\author{\normalsize Takeru K. Suzuki}
\address{\rm{School of Arts \& Sciences, University of Tokyo, 3-8-1, Komaba, Meguro, Tokyo 153-8902, Japan}}

\begin{abstract}
Recent theoretical studies suggest the existence of low-mass zero-metal stars in the current universe.  
In order to study the basic properties of the atmosphere of low-mass first stars, we perform one dimensional magnetohydrodynamical simulations for the heating of coronal loops on low-mass stars with various metallicities. 
While the simulated loops are heated up to $\geq 10^6$ K by the dissipation of Alfv\'{e}nic waves originating from the convective motion irrespectively of the metallicity,  the coronal properties sensitively depend on the metallicity. Lower-metal stars create hotter and denser coronae because the radiative cooling is suppressed.
The zero-metal star gives more than 40 times higher coronal density than the solar-metallicity counterpart, and as a result, the UV and X-ray fluxes from the loop are several times higher than those of the solar metallicity star. We also discuss the dependence of the coronal properties on the length of the simulated coronal loops.

\end{abstract}


\section{Introduction}
First stars have a significant impact on the cosmic history.
They were formed at the end of the dark ages and contributed to the cosmic reionization and the subsequent chemical evolution.
One of the main subjects among the remaining questions is the determination of the mass-scale of the first stars. 
It is an essential problem because the picture of the cosmic evolution is different depending on what kind of the mass distribution they had.
According to current common understanding, first stars are regarded to be more massive with the typical mass $\sim 100M_{\odot}$ because of the lack of metal coolants, where $M_{\odot}$ is the solar mass (Omukai \& Palla 2001; Bromm et al. 2002; Yoshida et al. 2006; O’Shea \& Norman 2007; Hirano et al. 2014).

However, recent cosmological simulations indicate that the protostellar disc around massive first stars fragments into clouds to form low-mass first stars (Clark et al. 2011; Greif et al. 2012; Machida \& Doi 2013; Susa et al. 2014; Chiaki et al. 2016). 
If the first stars with $M\leq 0.8 M_{\odot}$ were actually formed, they survive in the present universe owing to their long lifetime.
However, low-mass zero-metal stars have not been discovered yet, in spite of extensive observations of metal-poor stars (Christlieb et al. 2004; Aoki et al. 2006; Keller et al. 2014).   
Reasons of the non-detection have been discussed from various viewpoints of the surface pollution by heavy elements (Yoshii 1981; Komiya et al. 2015; Tanaka et al. 2017; Tanikawa et al. 2018)

Zero-metal stars with the mass $M_{\star} < 0.9 M_{\odot}$ have a convection zone (Richard et al. 2002), similarly to the Sun. 
The surface convective motion in the Sun triggers various magnetic activities, such as (micro- or nano-) flares (e.g. Parker 1988; Shimizu 1995), waves (e.g., Okamoto et al. 2007; McIntosh et al. 2011) and spicules (e.g. Suematsu et al. 1995; De Pontieu et al. 2007). These magnetic events play a role in uplifting the kinetic energy of the convection to the upper atmosphere, which leads to the formation of the hot corona $\gtrsim 10^6$ K (Grotrian 1939; Edl\'{e}n 1943) that emits X-rays and the acceleration of the solar wind (Parker 1958). Low-mass main-sequence stars possess a surface convection zone, and they also show the activity of coronae (Ribas et al. 2005) and stellar winds (Wood et al. 2005), which are believed to be originated from the magnetic activity.

The properties of the stellar coronae and winds are different for different stars. While the primary factor that controls the magnetic activity is known to be the stellar rotation, which decreases with the stellar age (e.g. Skumanich 1972; G\"{u}del 2007), the mass and metallicity are also regarded to be important ingredients. The metallicity determines the efficiency of the radiative cooling in the stellar atmosphere, and therefore, it directly affects the physical conditions of the corona. Suzuki (2018) investigated the heating and acceleration of the plasma gas by Alfv\'{e}n waves in an open magnetic flux tube by magnetohydrodynamical (MHD) simulations. Alfv\'{e}n waves hardly dissipate on account of the incompressible nature. Hence they can carry the kinetic energy a long distance and it goes into the thermal energy in the upper layer (Alfv\'{e}n 1947). 
He concluded that the stellar wind and the X-ray emission from the corona are stronger for lower metallicity, because the coronal density is higher owing to the inefficient cooling.

However, the X-rays are dominantly emitted from the coronal plasma in closed magnetic structures rather than from the plasma in open magnetic flux regions because the gas confined in closed magnetic loops is systematically denser than the gas streaming out freely of open regions. Therefore, when we argue the X-ray and UV emissions from stellar coronae, we should investigate the contribution from closed loops. The main purpose of the present paper is to reveal the properties of coronal loops in zero/low-metal stars and study their emitted radiation.

This paper consists of the following sections.
In Section 2, our loop model and MHD simulations are described.
In Section 3, we show the results of dynamical evolution, loop profiles and UV and X-ray fluxes for the stars with different metallicities. 
The dependence on the loop length is also presented.
Finally we summarize our study and discuss the treatments of our simulations and the future prospects in Section 4.
\\

\section{Method}
We perform 1-dimensional magnetohydrodynamic simulations for the heating of coronal loops by the dissipation of Alfv\'{e}n waves. The velocity perturbation, $\delta v$, which is driven by surface convective motion, is injected from the footpoints of a magnetic loop. It excites MHD waves that propagate upward.  In this study, we perform simulations with different stellar metallicities, $Z=1, 0.1, 0.01, 0.001, 0Z_{\odot}$, where $Z_{\odot}$ stands for the solar metallicity, and investigate the dependence of physical properties of the coronal loops on metallicity.

We numerically solve time-dependent MHD equations including radiative cooling and thermal conduction:

\begin{align}
 \frac{\partial \rho}{\partial t} + \nabla \cdot \rho\bm{v} = 0,
\end{align}
\begin{align}
\begin{split}
 \rho\frac{\partial \bm{v}}{\partial t} = &-\nabla \left(P + \frac{B^2}{8\pi}\right)+ \frac{1}{4\pi} \left (\bm{B} \cdot\nabla\right) \bm{B}  \\
&- \rho\left(\bm{v} \cdot \nabla\right)\bm{v}- \frac{\rho GM}{R^2} \hat{R},
  \end{split}
\end{align}
\begin{align}
 \frac{\partial \bm{B}}{\partial t} = \nabla\times\left(\bm{v} \times \bm{B}\right) 
\end{align}
and
\begin{align}
\begin{split}
 &\rho\frac{d}{dt}\left(e+\frac{v^2}{2}+\frac{B^2}{8\pi\rho}- \frac{GM}{R^2}\right)  \\
 &+\nabla \cdot \left[\left(P+\frac{B^2}{8\pi}\right)v-\frac{B}{4\pi}\left(B\cdot\bm{v}\right)\right]  \\
 &+ \nabla\cdot F_c  +  q_R = 0   
 \end{split}
\end{align}
In the above equations, $R$ is a distance from the center of a star and the unit vector is defined as $\hat{R}=\bm{R}/R$. $G$ and $e$ denote the gravitational constant and the internal energy per mass which satisfies the relation $e=\frac{P}{(\gamma - 1)\rho}$ where $\gamma=5/3$ is the ratio of specific heats. $F_c$ is the thermal conduction for fully ionized plasma with the form  $F_c=\kappa T^{5/2}\frac{\partial T}{\partial s} $  where $\kappa$ is the Spitzer conductivity and $q_R$ is radiative cooling rate, which is described in Section 2.3. 

\begin{figure}[t]
 \centering
\includegraphics[width=8cm]{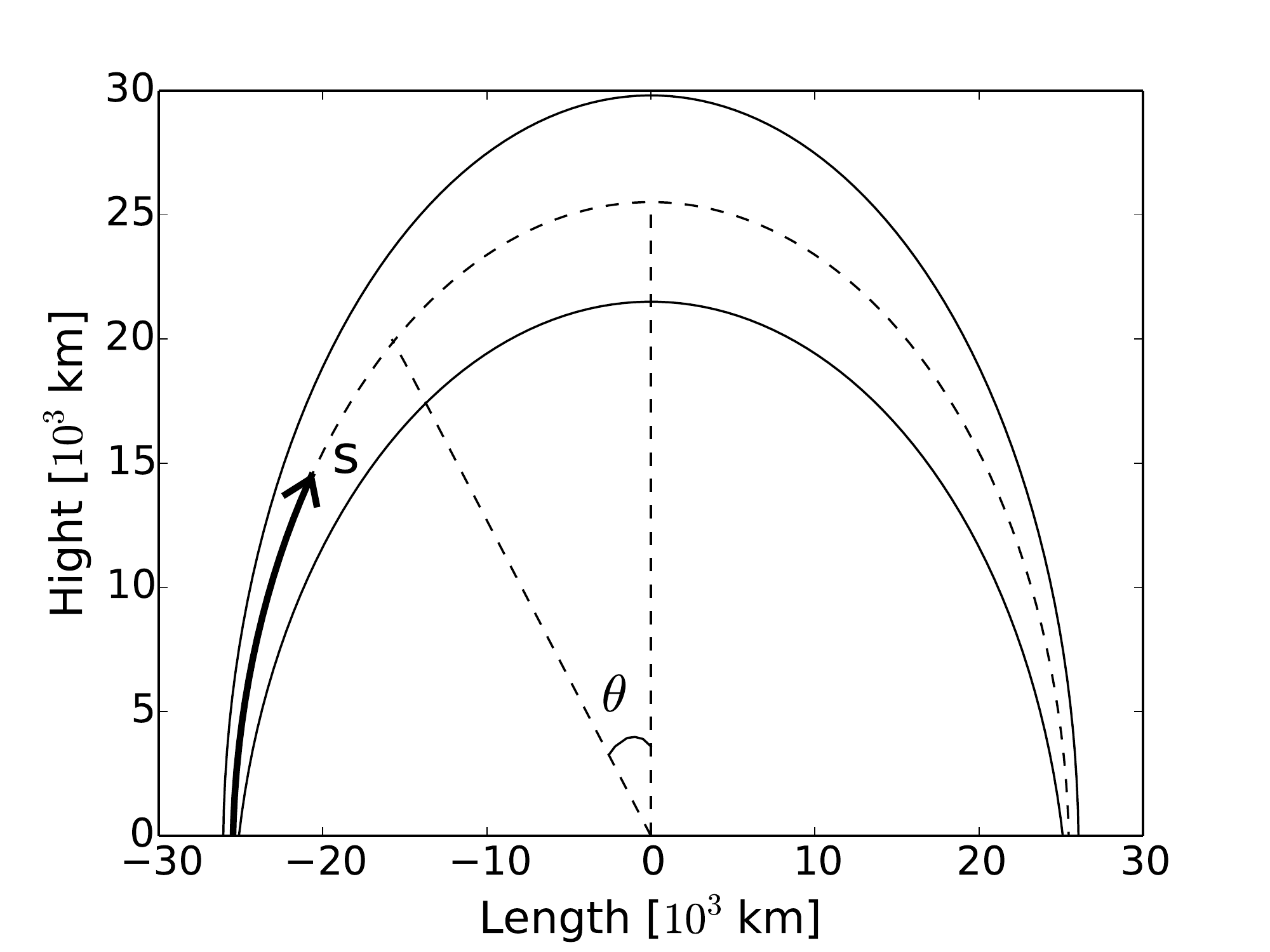}
\caption{Geometry of our coronal loop model.
The cross-sectional area expands 200 times that at the footpoints. The calculation is done along s direction (dashed line).}
\end{figure}

\subsection{Set up}
We adopt a magnetic loop, of which the geometrical configuration is similar to that of Moriyasu et al. (2004). The loop is considered as a semicircular magnetic flux tube of $l$ in length and both ends are rooted in the photosphere (Figure 1). We adopt the loop length $l = 0.8\times 10^5$ km for the standard case and $l = 1.1\times 10^5$ km  and $1.6\times 10^5$ km for other cases (mainly discussed in Section 3.3). We perform magnetohydrodynamical simulations in a one dimensional loop that is along a $s$ coordinate. Therefore in solving MHD equations described before, the gravitational term is replaced by $ -\frac{\rho GM}{R^2}\hat R \sin{\theta}$ where the angle $\theta$  is taken from the loop top (Figure 1).   
 Furthermore in order to handle the torsional motions, the velocity and magnetic fields have three components, $\bm{v}= (v_{s},v_{\perp1},v_{\perp2})$, $\bm{B}= (B_{s},B_{\perp1},B_{\perp2})$, where the latter two components have the directions perpendicular to $s$, however  
eqs.(1)-(4) should satisfy the relation $\frac{\partial }{\partial \perp 1} = 0$ and $\frac{\partial}{\partial \perp 2} = 0$. This treatment is so-called 1.5D MHD calculation.

The cross-sectional area at the loop top expands 200 times that of the footpoints and we define a loop expansion factor,

\begin{align}
\begin{split}
f(s) = 200 \times \frac{1}{2} [ \rm{tanh} \{\it{a}(\frac{-|s|/R+b}{h/R}+\frac{\rm \pi}{\rm4})\} + \rm1],
\end{split}
\end{align}
which has the relation $B_s = B_{\rm ph}/f(s)$.  $a = 6.86$ and $b = 0.02$ are free parameters that determine the shape of $f(s)$. We note that it is necessary to take into account $f(s)$ when we calculate the divergence and rotation of a vector, $\bm{X}$: 
\begin{align}
&\nabla \cdot \bm{X} = \frac{1}{f}\frac{\partial}{\partial s} fX_s  
\end{align}
and
\begin{align}
 &\nabla \times \bm{X} = \frac{1}{\sqrt{f}}\frac{\partial}{\partial s} \sqrt{f}X_s,
 \end{align}
where $X_s$ is the $s$ component of $\bm{X}$.

\subsection{Parameters}

The stellar mass is set at 0.8$M_{\odot}$ and  the effective temperature is $T_{\rm eff}=5100$ K. 
We adopt the stellar radius $R=5.13 \times 10^5$km for $M=0.8M_{\odot}$ with the solar metallicity (Yi et al. 2001, 2003). 
At the photosphere, the density is $\rho_{\rm ph}=4.37\times 10^{-7}  \rm g/\rm cm^3$, the magnetic field strength is $B_{\rm ph}=1.98$ kG and the velocity perturbation is $\delta v = 1.0$ km/s in all three directions. In this paper, we adopt the same values for all the stars with different metallicities. We inject the velocity perturbations with the spectral shape $\propto \omega^{-1}$ with the frequency $\omega$ between $\omega_{\rm max}^{-1} \approx $ 20 $\rm s$ and $\omega_{\rm min}^{-1} \approx$ 1800 $\rm s$ roughly same as the solar case. 

We set the initial temperature, $T_{\rm init} = T_{\rm eff}$, in the entire loop.
We set up the initial density that is higher than that is expected from the hydrostatic equilibrium with $T_{\rm init} = T_{\rm eff}$ to avoid the severe CFL (Courant-Friedrichs-Lewy) condition; if we adopted the hydrostatic density with $T_{\rm init}=T_{\rm eff}$, the density at the loop top was so low that the Alfv\'{e}n speed there was too fast.
After the simulation starts, the gas near the loop top falls down because the density is higher than the hydrostatic value. However, the falling gas is  eventually pushed back upward by the upgoing Alfv\'{e}n waves from below. Therefore, the artificially high initial density does not affect the result after the high-temperature corona is formed.
 
The numerical scheme is 2nd Godunov-MOC (Method-of-Characteristics; Stone \& Norman 1992) method (Sano \& Miyama 1999; Suzuki $\&$ Inutsuka 2005).
We take the grid size $\delta s=10$km for the chromospheric region and $\delta s=100$ km for the coronal region that can resolve the shortest wavelength of the injected waves by 10 grid points. 

\subsection{Radiative Cooling}
We explicitly take into account the metallicity dependence of the radiative cooling,  and adopt different prescriptions below and above the bottom of the transition region that divides the chromosphere and the corona. 

In the optically thin region with T $\geq 10^4$K, the cooling occurs through the collision between an electron and an ion, therefore the cooling rate $q_R$ [erg $\rm cm^{-3} \rm s^{-1}$] is calculated by;  
\begin{align}
& q_R = \Lambda n n_e
\end{align}
$\Lambda$ is the cooling function that depends on metallicity (Figure 2.) referred from Sutherland \& Dopita(1993). $n$ is ion number density and  $n_e$ is electron number density.

In the optically thick region with T $\leq 10^4$ K, we use the empirical formula introduced  in Suzuki(2018); 
\begin{align}
& q_R = 4.5 \times 10^9 \rho \left(0.2 + 0.8 Z/Z_{\odot} \right)
\end{align}
This is an extension  from the empirical cooling rate of the solar chromosphere $q_R = 4.5 \times 10^9 \rho$ (Anderson $\&$ Athay 1989); we utilized the fact that H$\alpha$ emission contributes about 20$\%$ to the total emission in the solar chromospheric region and that the rest 80\% is contributed from heavy elements.\\

\begin{figure}[t]
  \centering
\includegraphics[width=8cm]{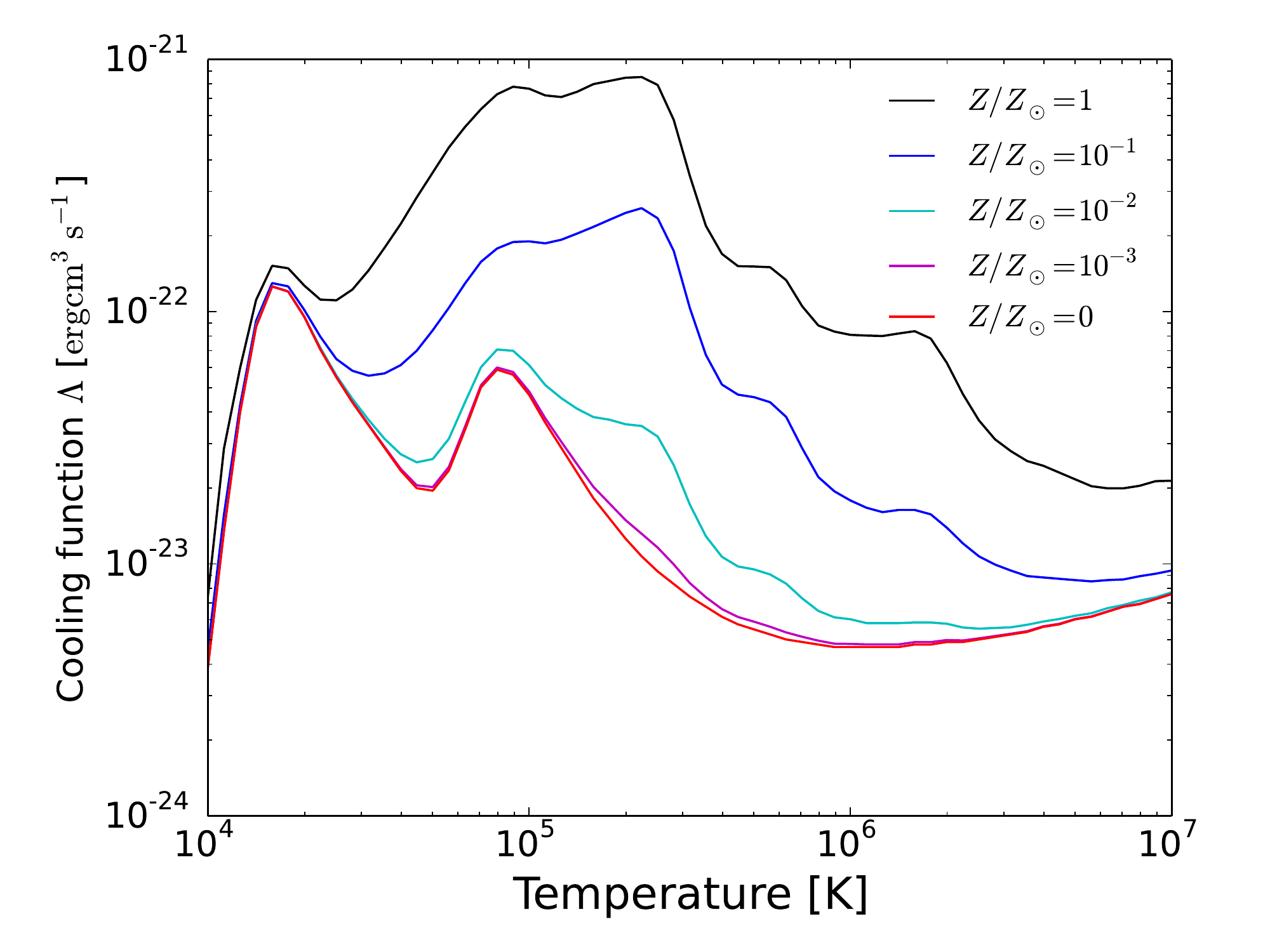}
\caption{Cooling function for different metallicities (Sutherland \& Dopita1993).}
\end{figure}

\section{Results}
\subsection{Dynamical Evolution}
In our simulations, the corona is heated by the dissipation of Alfv\'{e}nic waves by the nonlinear mode conversion to compressible waves (Kudoh \& Shibata 1999; Suzuki \& Inutsuka 2005, 2006); the qualitative aspect of the wave dissipation does not depend on the metallicity and loop length.
The Alfv\'{e}nic waves that are excited from the surface convection and propagate upward are in part converted to longitudinal slow MHD waves by the nonlinear mode conversion; fluctuations of magnetic pressure associated with Alfv\'{e}nic waves excite longitudinal density perturbations (Hollweg 1982); slow MHD waves are also generated from Alfv\'{e}n waves via parametric decay instability (Goldstein 1978; Terasawa et al.1986; Shoda et al. 2018).  

\begin{figure}[t]
  \centering
\includegraphics[width=9cm]{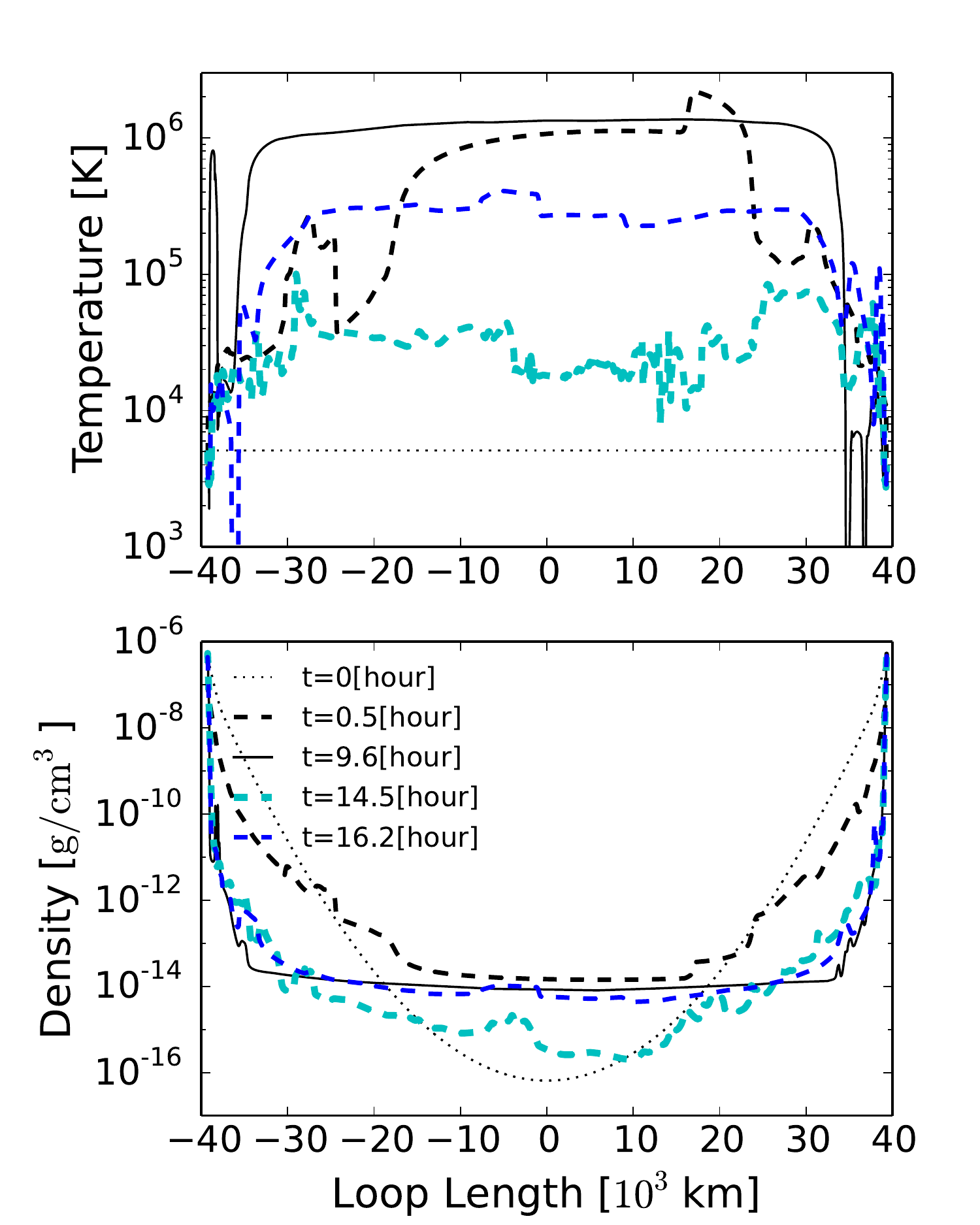}
\caption{Time evolution of the loop temperature (top) and density (bottom) with $Z=Z_{\odot}$. The profiles are plotted for t=0 hours (dotted line), t=0.5 hours (black dashed line), t=9.6 hours (black solid line), t=14.5 hours (cyan dashed line) and t=16.2 hours (blue dashed line).}
\end{figure}

 \begin{figure}[t]
  \centering
\includegraphics[width=9cm]{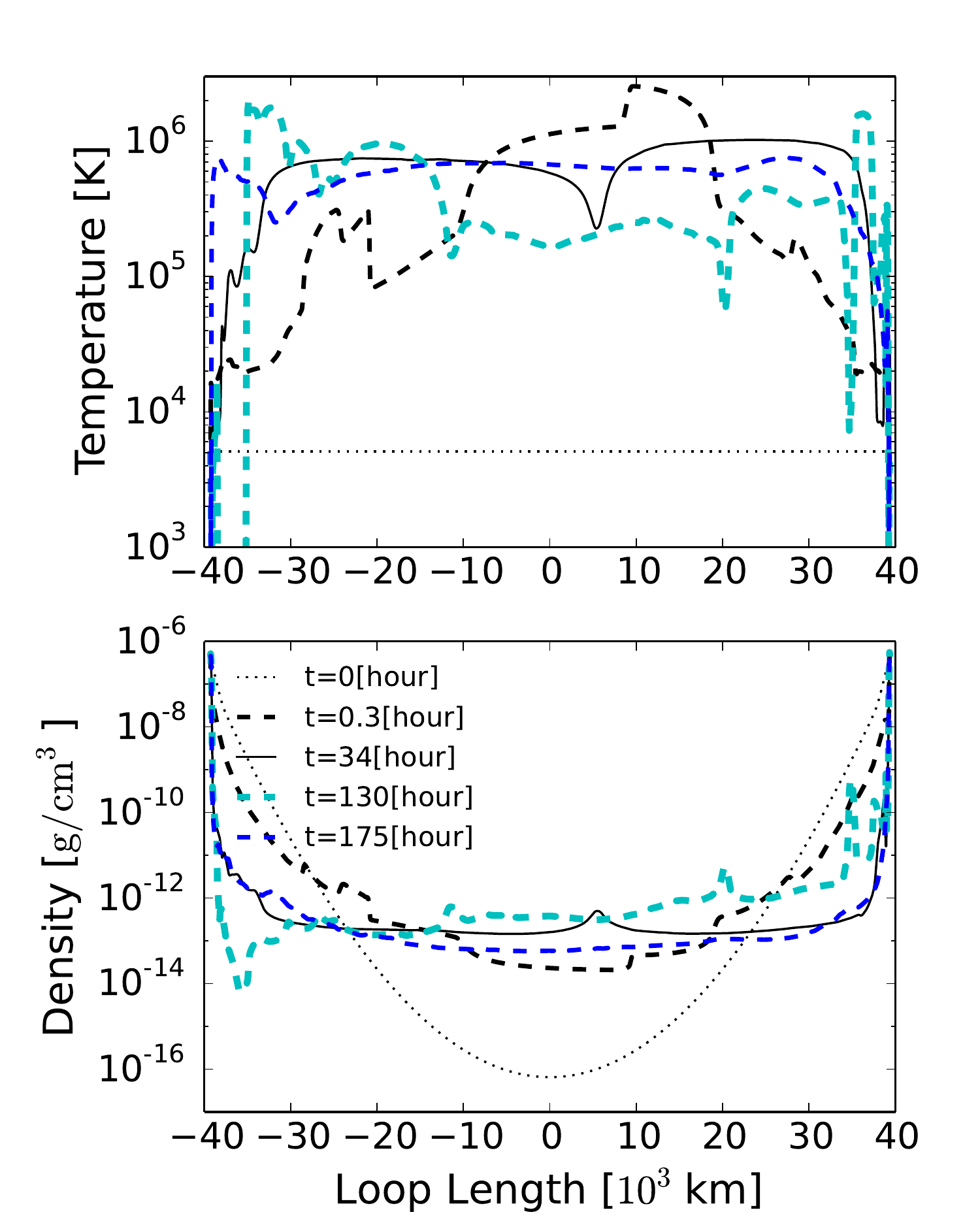}
\caption{Time evolution of the loop temperature (top) and density (bottom) with $Z=0$. The profiles are plotted for t=0 hours (dotted line), t=0.3 hours (black dashed line), t= 34 hours (black solid line), t= 130 hours (cyan dashed line) and  t=175 hours (blue dashed line).}
\end{figure}

Our results in the bottom panels of Figure 3 \& 4 which show the time evolution of the loop density with $Z=Z_{\odot}$ and $Z=0$ indicate that  the density perturbations are actually generated and the longitudinal MHD waves are excited after the simulation starts. These compressive waves are amplified through the upward propagation in the stratified atmosphere.  They eventually steepen to form shocks, which heats up the hot corona with $T>10^6$K. 
However, in realistic 3D simulations, the characteristics of the wave dissipation may be different from our results and we will discuss this issue in Section 4.

\begin{figure*}[t]
  \centering
\includegraphics[width=18cm,height=4cm]{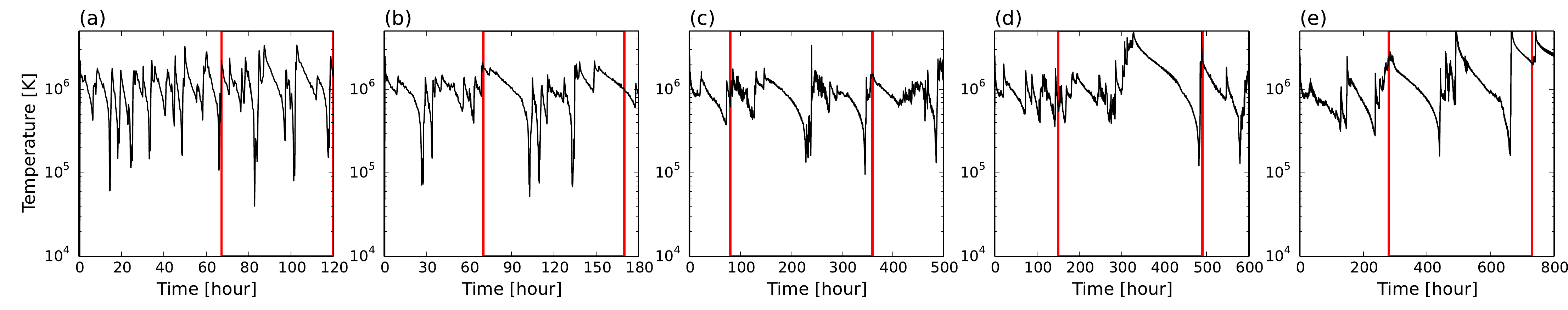}
\caption{Time variation of the temperature at loop top. \\
(a)$Z/Z_{\odot}=1$ (b)$Z/Z_{\odot}=0.1$ (c)$Z/Z_{\odot}=0.01$ (d)$Z/Z_{\odot}=0.001$ (e)$Z/Z_{\odot}=0$}
\end{figure*}

\subsubsection{$Z=Z_{\odot}$}

Figure 3 shows the time evolution of the loop temperature with $Z=Z_{\odot}$. From the initial cool state (dotted line), the loop is instantly heated (dashed line) by the dissipation of Alfv\'{e}nic waves. After $\approx$ 9.6 hours, the quasi-steady corona is formed; the temperature distribution is almost flat with $T\approx 10^6$ K by thermal conduction (solid line). Although this quasi-steady corona is sustained for several hours, the temperature around the loop top eventually drops to $10^4-10^5$ K (cyan dashed line). After the corona disappears for $\approx$ 1-2 hours, the temperature is again rapidly heated up (blue dashed line) and the quasi-steady corona is recovered. The formation and destruction of the corona repeatedly occurs afterward in a quasi-periodic manner.

The cyclic evolution of coronal loops has been investigated by previous studies (Kuin $\&$ Martens 1982; M$\ddot{\rm u}$ller et al. 2003,2004; Mendoza-Brice$\tilde{\rm n}$o et al. 2005). The mechanism of this cyclic behavior can be explained as follows;  
when the quasi-steady corona is formed, the chromosphere is gradually heated up by the downward thermal conduction. Consequently, the chromospheric material evaporates into the corona. The coronal density gradually gets higher as a result of the mass supply by the chromospheric evaporation.  The increase of the density, however, enhances the radiative cooling, which causes the decrease of the temperature.  In addition, the temperature range of $10^5$ K $\leq T \leq 10^6$ K is thermally unstable. Therefore, once the coronal temperature, $\geq 10^6$ K, cannot be sustained, it suddenly drops below $T<10^5$ K.  Accordingly, the pressure, which is proportional to the temperature, also drops so that it cannot support the coronal gas against the downward gravity. The coronal material mostly falls down to the chromosphere. Once the density in the upper region drops, the radiative cooling is suppressed.  As a result, the temperature rapidly rises again so that the quasi-steady corona is revived. These consecutive processes occur repeatedly in a quasi-periodic manner (Figure 5). 
 
\subsubsection{$Z=0$}
The time evolution of the case with $Z=0$ is shown in Figure 4. The gas is heated up to $T\approx 10^6$ K at $t=0.3$ hour (dashed line) from the initial cool atmosphere (dotted line). After the relatively cool corona is kept for a while (solid line), the temperature drops to $T\approx (2-3) \times 10^5$ K (cyan dashed line)  and again,  rises to $T\geq 10^6$ K (blue dashed line). In the zero metallicity case, the time for a cycle is quite long even though the qualitative behavior of the evolution is similar to that of the solar metallicity case (Figure 5).

\subsubsection{Dynamical Evolution -Summary-}
Figure 5 shows the time variation of the temperature at the loop top of all the different metallicity cases we simulated. As discussed previously, the case with $Z=Z_{\odot}$ shows that the formation and destruction of the corona take place iteratively. Although the variation of the peak temperature is not purely periodic, the rough average period is $\approx$(8-10) hours. The iterative cycle is also observed in all other metallicity cases. However, the average period is longer for lower metallicity, especially it is an order of 100 hours for $Z=0$.

 The negative dependence of the average period on metallicity can be interpreted by the cooling time that strongly depends on metallicity. The cooling time is defined as $\tau_{\rm cool} \approx \frac{\rho e}{q_{\rm R}}$. $\tau_{\rm cool}$ is longer for lower $Z$ because $q_{\rm R}$ is smaller (Figure 2). Therefore, it takes longer time for lower metallicity stars before the collapse of the corona occurs by the enhanced cooling. 
 
\subsection{Comparison of Time-averaged Profiles}
\subsubsection{Loop Profiles}
We compare the time-averaged loop structures of different metallicities. The average is taken during the period shown by the red frame in Figure 5. The durations of the averages are set up in order that at least multiple ($\approx 6$ for $Z=Z_{\odot}$ and $\approx$ 2 for $Z=0$) cycles are covered. 

Figure 6 compares the temperature and density distributions of all the simulated cases with the five different metallicities. Both temperature and density are higher in lower metal stars. The profiles of the case with $Z=0.001 Z_{\odot}$ is almost the same as those of the zero metal case, which indicates that the coronal properties of a star with $Z\leq0.001Z_{\odot}$ are not different from those of a zero metal star even though it has a finite amount of heavy elements. 

\begin{figure}[t]
\begin{centering}
\includegraphics[width=80mm]{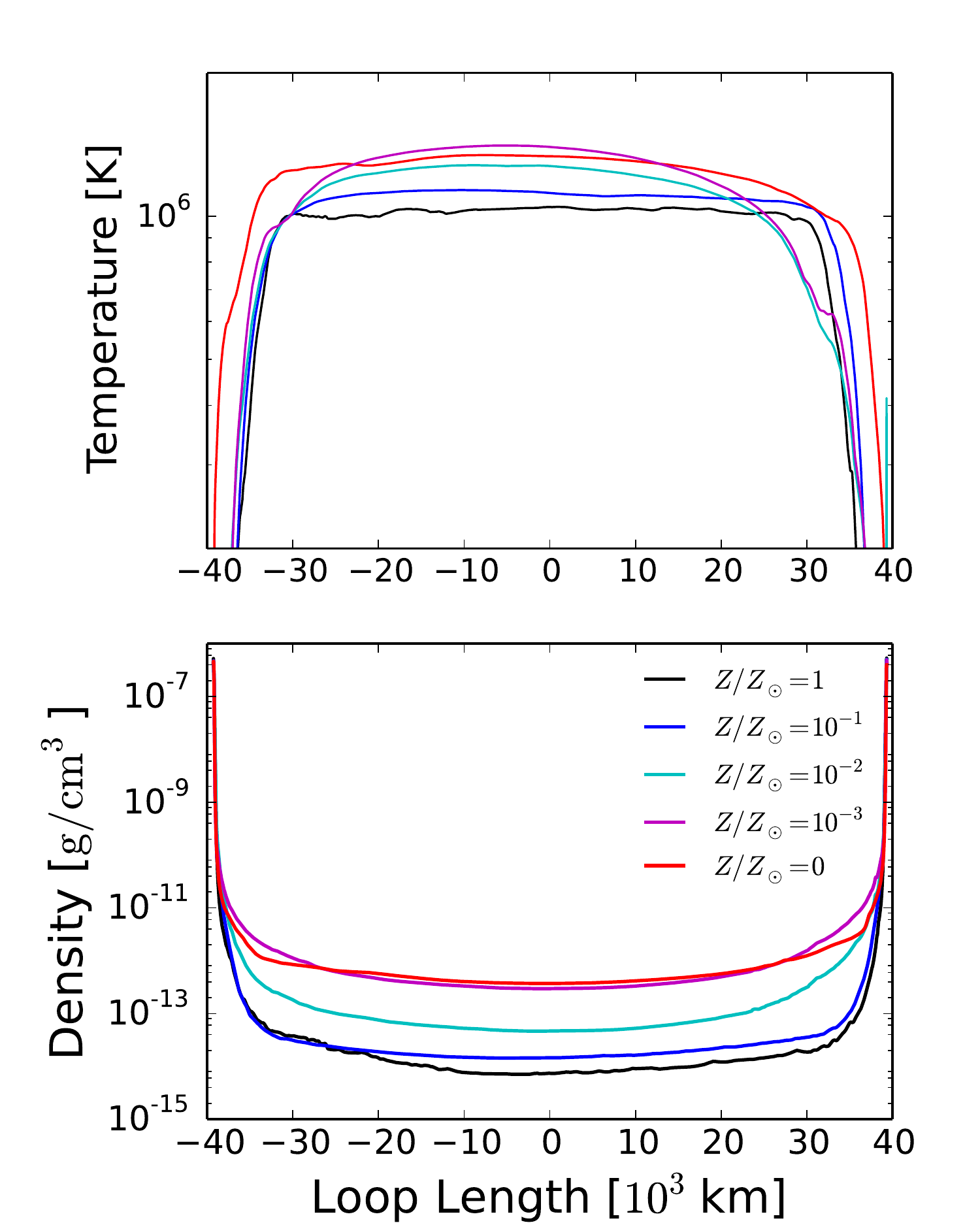}
\end{centering}
  \caption{Time-averaged temperature (top) and density (bottom) of the different metallicity cases. }
  \label{fig:3}
  \end{figure}

The temperature and the density in the coronal region of the zero metal star is $\approx 1.5$ times and $\approx 40$ times higher respectively than those of the solar metal star. The suppression of the radiative cooling is a part of the reason why both coronal temperature and density have the negative dependences on $Z$. In addition to this cooling effect, the following positive feedback involving the chromospheric evaporation and the reflection of waves causes the very dense corona of the zero and very low metallicity stars. The higher temperature leads to more efficient chromospheric evaporation, which forms denser corona. The difference between the chromospheric density and the coronal density gets smaller for lower metallicity stars. As a results, a larger fraction of the Alfv\'{e}n waves that are excited from the photosphere can be transmitted to the corona, because of the suppression of the reflection (Verdini et al. 2012; Suzuki 2018). The larger transmissivity of the waves further gives larger heating by the wave dissipation.  These consecutive processes operate as a positive feedback, and hence, the coronal density of the zero metal star is much larger than that of the solar metallicity star.

\subsubsection{Heating Rates}

In order to understand the coronal heating for different metallicity stars, we compare the heating rates as a function of distance along the loop.
The dissipation rate $(\rm{erg}/cm^3 s)$ of Alfv\'{e}nic waves can be written as 
\begin{align}
&Q(s) = \nabla \cdot \{- \frac{1}{4\pi}B_s(\bm{ B_{\perp}\cdot v_{\perp}}) + v_s(\frac{1}{2}\rho v_{\perp}^2 + \frac{B_{\perp}^2}{4\pi})\},
\end{align}
which corresponds to the heating rate of the surrounding gas. 
The first term represents the heating by Alfv\'enic waves and
the second term is due to the heat transfer by the advection. We note that the contribution from the second term is generally much smaller than that from the first term.

Figure 7 presents $Q(s)$ for the different metallicity cases. When the steady-state condition is achieved, the heating in the corona is balanced with the energy loss from the corona; $Q(s)$ is equal to the sum of the radiation from the corona and the downward thermal conduction to the chromosphere. The downward conductive flux is finally converted to the radiation from the transition region and the chromosphere, and therefore, the integrated $Q(s)$ along $s$ determines the total radiation from the simulated loops. 

In all the cases, the heating rate in the corona sharply decreases from much larger heating rate in the chromosphere. This is because the density in the corona is several orders of magnitude smaller than that in the chromosphere and the smaller heating rate is sufficient to maintain the hot corona. Turning to the metallicity dependence, the heating rate in the corona is higher for lower metallicity. This is because a larger fraction of the injected Alfv\'{e}nic waves is transmitted to the corona owing to the suppressed reflection (Section 3.2.1). 

 \begin{figure}[t]
\begin{centering}
\includegraphics[width=80mm]{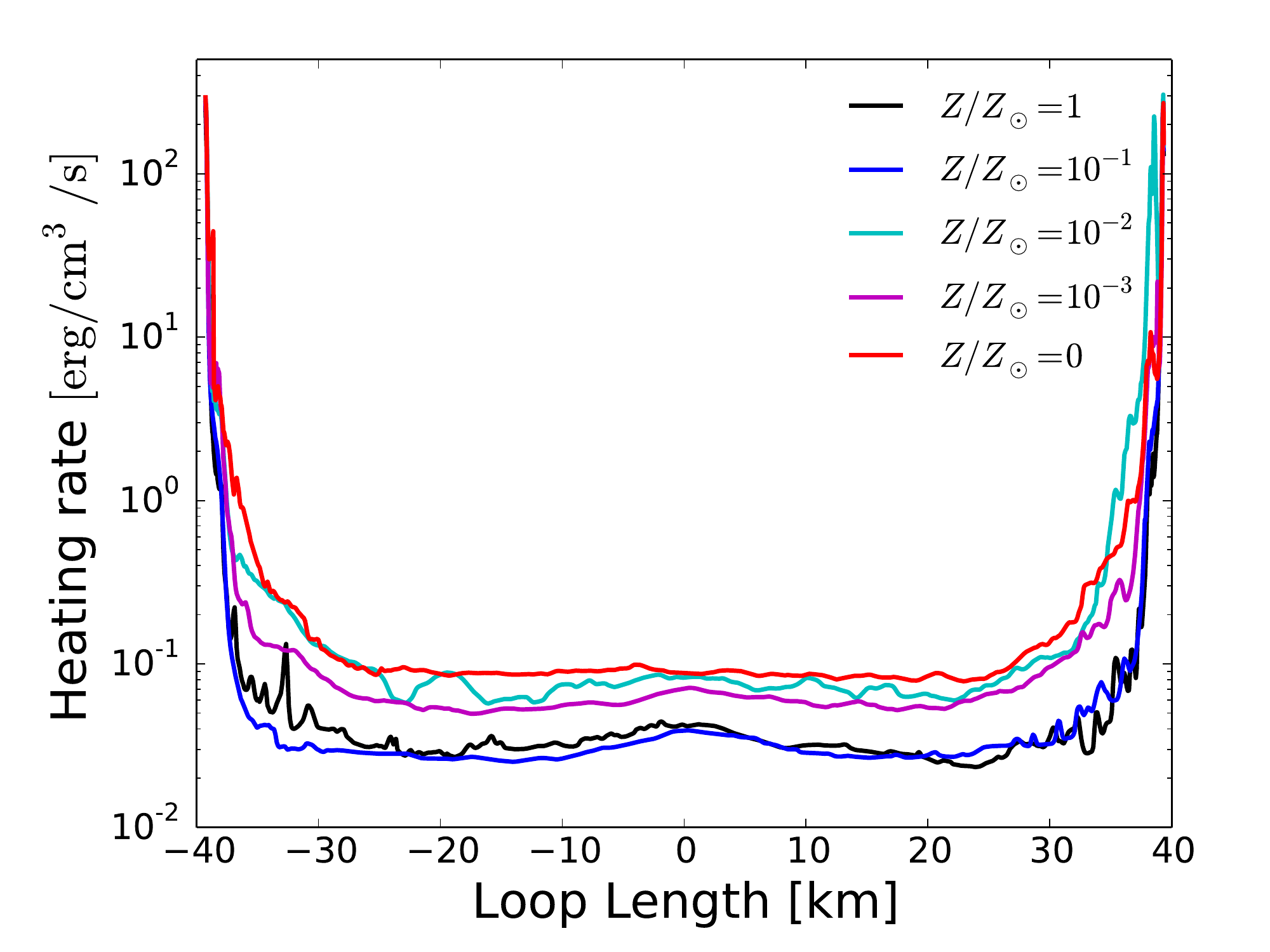}
\end{centering}
  \caption{Heating rates of the different metallicity cases.}
    \end{figure}    
      
\subsubsection{Differential Emission Measure Analysis}
In deriving the thermal properties of the simulated loops, differential emission measure, 
\begin{align}
&\mathrm{DEM}(T) = n_{e}^2 \frac{ds}{dT},
\end {align}
 is a useful indicator that characterizes the temperature distribution of gas and determines X-ray and EUV radiation, which we will discuss later. 
Figure 8 shows the comparison of DEM$(T)$  for the different metallicity cases.
It satisfies the condition of the constant pressure equilibrium which has the relation DEM$(T) \propto T^{-3}$(Athay 1976).
The corona is denser for lower metallicity so that DEM$(T) \propto n_e^2$ takes much larger values. 

  \begin{figure}[t]
\begin{centering}
\includegraphics[width=80mm]{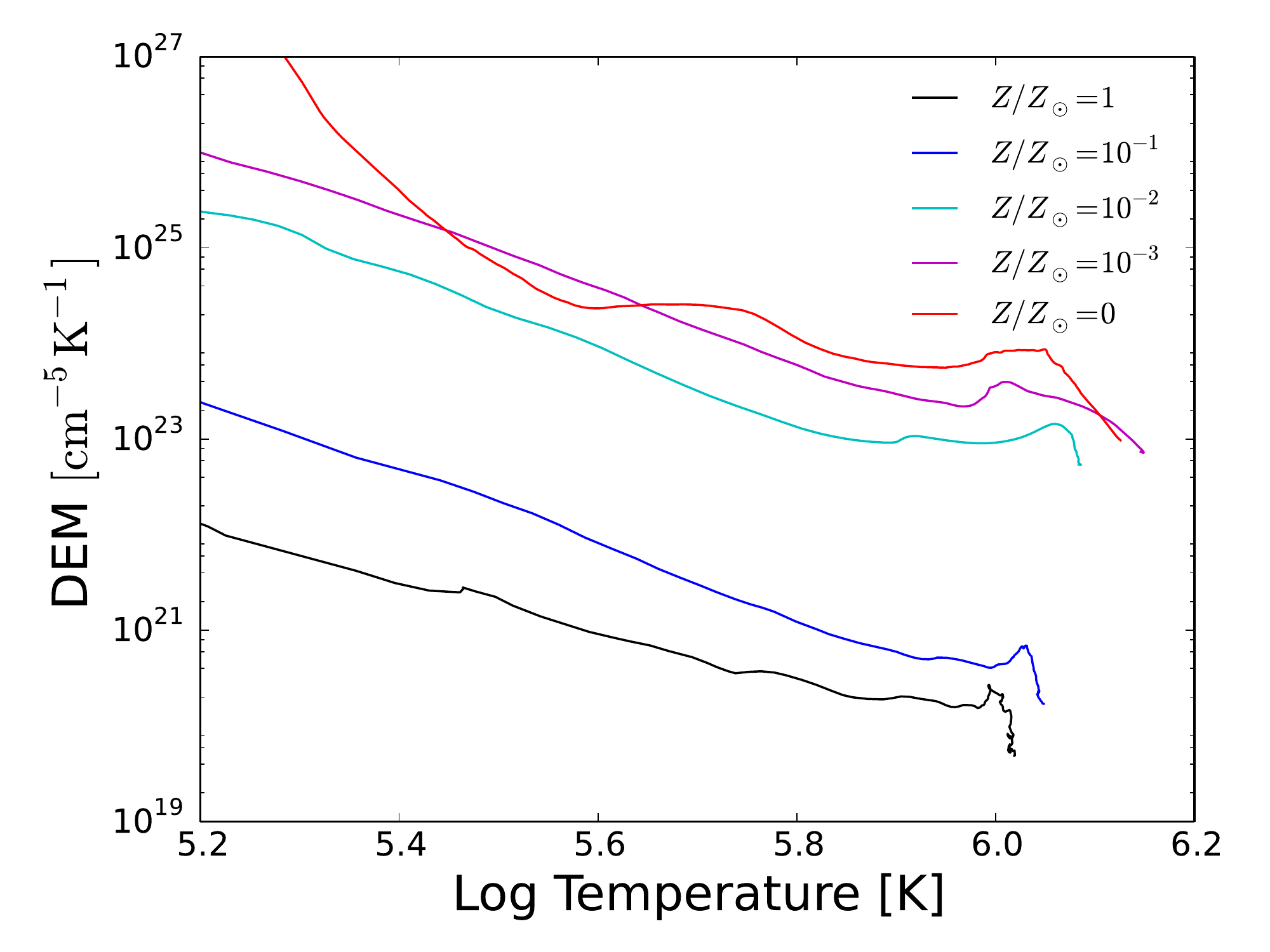}
\end{centering}
  \caption{DEM distributions for the different metallicity cases.}
    \end{figure}

\subsubsection{UV and X-ray Fluxes}

We examine the radiative fluxes emitted from  the simulated loops. We separate them into the component from the transition region with $2\times 10^4$ K $\le T \le 5\times 10^5$ K and the component from  the corona with $5\times 10^5$ K $<T$; we call the former UV and the latter (soft) X(-rays). 

In Figure 9, we present these two components
\begin{align}
&L_{UV} = \frac{1}{f_{\rm top} } \times 4 \pi R^2 \int_{s( T \geq 2\times10^4 K)}^{s(T\le 5\times 10^5{\rm K})} \Lambda n n_e f(s) ds 
\end{align}
\begin{align}
&L_X =  \frac{1}{f_{\rm top} } \times 4 \pi R^2 \int_{s(T > 5 \times 10^5 K)} \Lambda n n_e f(s) ds  
\end{align}
in units of luminosity, erg s$^{-1}$. $f_{\rm top} = 200$ is the value of $f(s)$ at the loop top. Since our simulations are done in 1D flux tubes, we can only calculate radiation flux. In order to estimate the radiation luminosity, we assume that the total surface area is covered by $(4\pi R^2/ A_{\rm top})$ loops, where $A_{\rm top} =f_{\rm top}\times$ (cross sectional area of a single loop) is the cross sectional area of the flux tube at the loop top. 

For $l=0.8\times 10^5$ km (solid lines), both $L_{\rm X}+L_{\rm UV}$ and $L_{\rm X}$ roughly increase  with decreasing $Z$. In the case of $Z = 0$, $L_{\rm X} + L_{\rm UV}$ is $\approx 4.1$ times and $L_{\rm X}$ is $\approx 3.1$ times larger than that of  $Z=Z_{\odot}$. On the other hand, although the rough tendency between $L$ and $Z$ is anti-correlation, the detailed trend is not monotonic. For example, $L_{\rm X}$ decreases for decreasing $Z$ from $Z_{\odot}$ to $0.1Z_{\odot}$. $L$ is determined by the cooling function, $\Lambda$, multiplied by the DEM (eqs.8 \& 11). This exceptional positive correlation arises because the rapid decrease of $\Lambda$ on  $Z$ cannot be compensated by the increase of the DEM. In contrast, $L_{\rm X}$ increases on decreasing $Z$ in the lower $Z(<0.01Z_{\odot})$ range, because the increase of the DEM dominates the decrease of $\Lambda$.

 \begin{figure}[t]
\begin{centering}
\includegraphics[width=80mm]{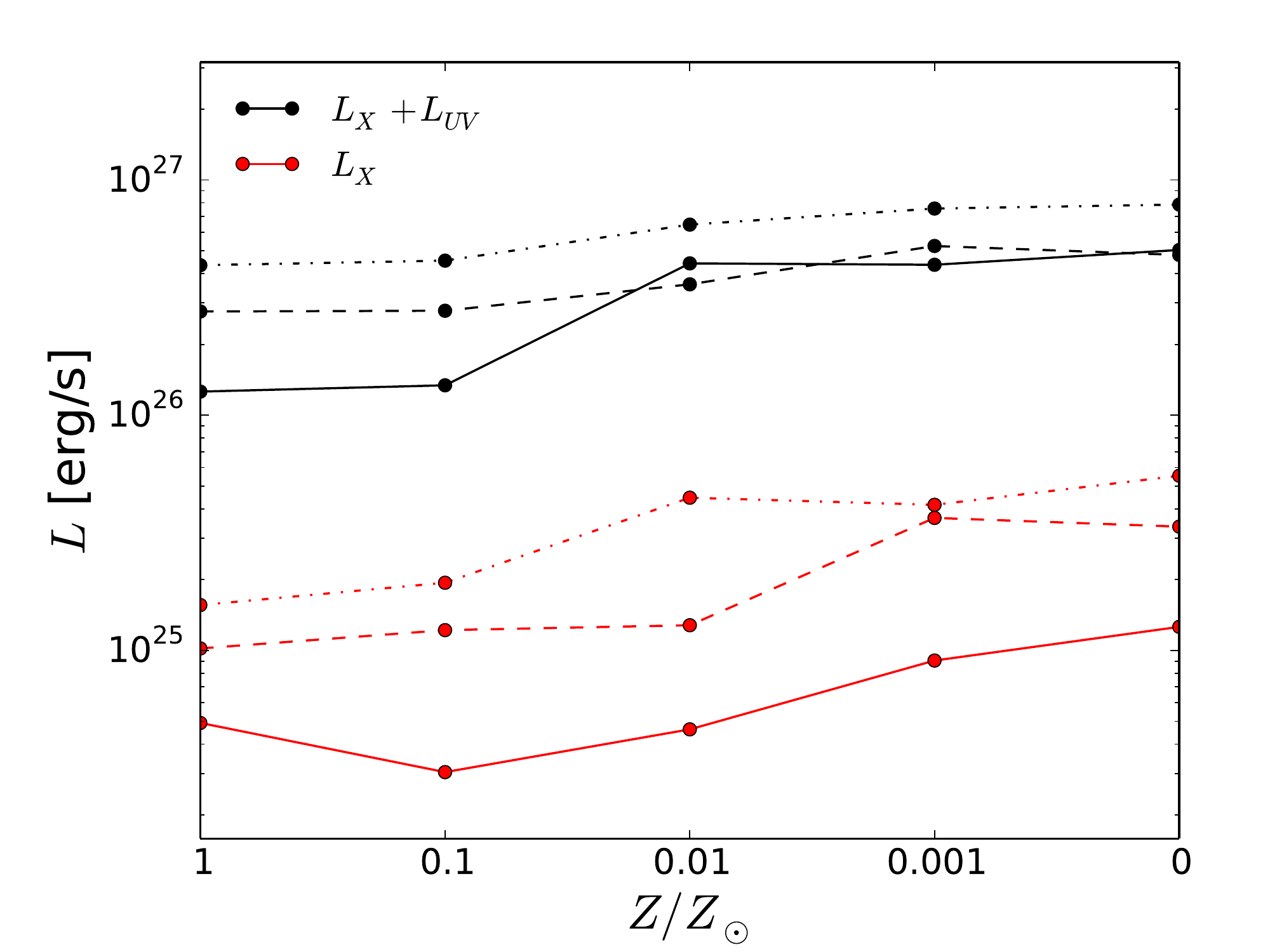}
\end{centering}
  \caption{Comparison of $L_{\rm X} + L_{\rm UV}$ (black lines) and $L_{\rm X}$ (red lines) for different metallicities. Line types corresponds to loop lengths: $l=0.8\times 10^5$ km (solid lines), $l=1.1\times 10^5$ km (dashed lines) and $l=1.6\times 10^5$ km (dashed-dot lines).}
  \label{fig:4}
  \end{figure} 
  
\subsection{Dependence on the Loop Length}
We carry out the same sets of simulations for different loop lengths, $l=1.1\times 10^5, 1.6\times 10^5$ km to investigate the dependence of the coronal structure on the loop length.

\subsubsection{Dynamical Evolution}
Figure 10 shows the time variation of the temperature at the loop top of the different metallicity cases for $l=1.6\times 10^5$ km. 
All the cases show the cyclic evolution and the negative dependence of the average cycle period on metallicity, which is the qualitatively same as the cases with the shorter $l=0.8\times 10^5$km (Figure 5 \& Section 3.1.3).  However, compared with Figure 5, the average period is longer for the longer loop with the same metallicity. In the case of $Z=Z_{\odot}$, the period is $\geq 10$ hours for $l=1.6\times 10^{5}$ km, while it is $\sim (8-10)$ hours for $l = 0.8\times 10^5$ km. In the lower metallicity cases with $Z \leq 0.001Z_{\odot}$, the period is further longer $>150$ hours and the loops evolve quite slowly.
The density at the loop top is lower for longer loops. Therefore, it is easier to sustain the hotter corona because the radiative cooling is less efficient (Figures 11 \& 12). 

\begin{figure*}[t]
  \centering
\includegraphics[width=18cm,height=4cm]{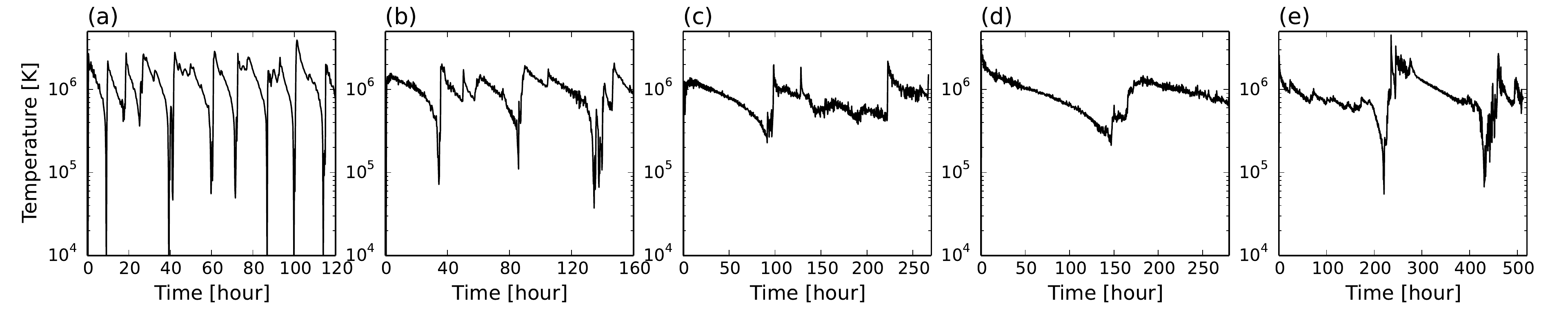}
\caption{Same as Figure 5 but for $l=1.6\times 10^5$km.}
\end{figure*}
 \begin{figure}[t]
\begin{centering}
\includegraphics[width=80mm]{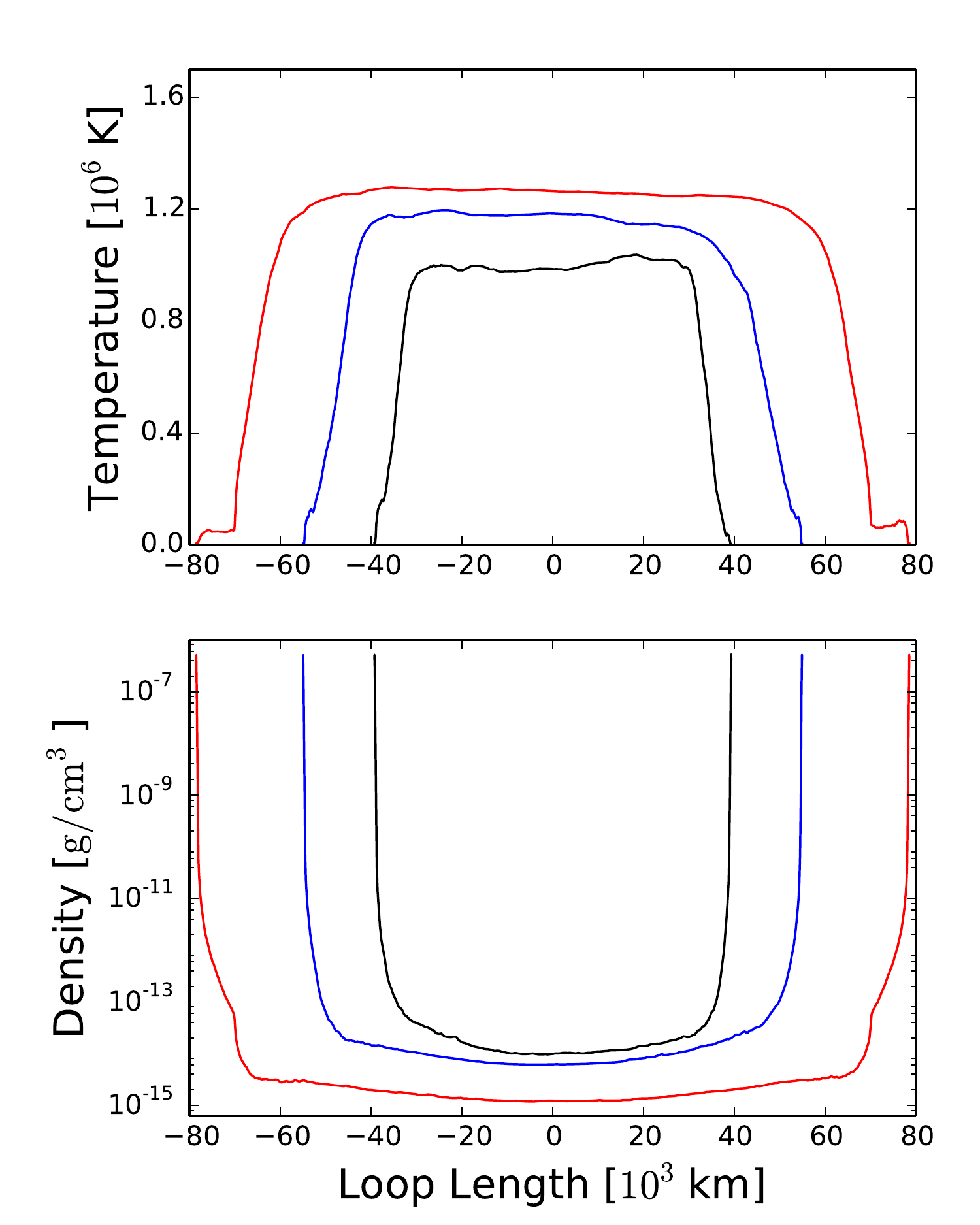}
\end{centering}
  \caption{Time-averaged loop structures of the different loop lengths with $Z=Z_{\odot}$. The length of each loop is $0.8\times 10^5$ km (black line), $1.1\times 10^5$ km (blue line) and $1.6\times 10^5$ km (red line). }
  \end{figure}
   \begin{figure}[t]
\begin{centering}
\includegraphics[width=80mm]{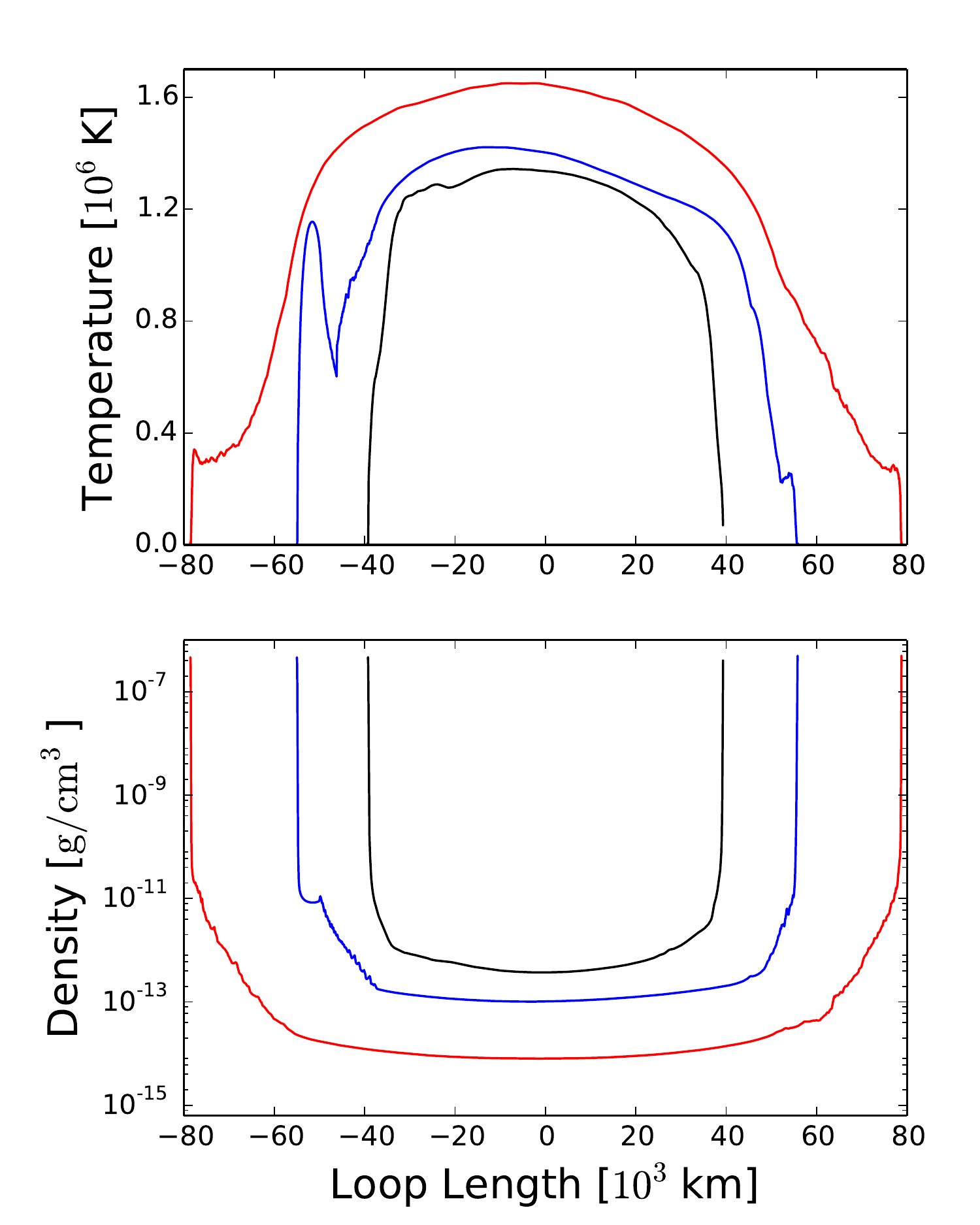}
\end{centering}
  \caption{Same as Figure 11 but for $Z=0$.}
  \end{figure}

\subsubsection{Profiles of the Different Loop Length}
We compare the time-averaged temperature and density distributions for different loop lengths in Figures 11 ($Z=Z_{\odot}$) and 12 ($Z=0$), respectively. Both cases show that the physical properties of the loops depend on the loop length; longer loops give higher temperatures and lower densities than shorter loops with the same metallicity. The loop top is located at a higher altitude for a longer loop, which gives the lower density there for the nearly hydrostatic structure. Therefore the gas in the corona can be heated to higher temperatures due to the suppressed cooling in the lower density condition even with the same metallicity.


\subsubsection{$L_{\rm X}$ and $L_{\rm UV}$}
The radiative luminosities $L_{\rm X}$ and $L_{\rm UV}$ are shown in Figure 9 for $l=1.1\times 10^5$ km (dashed lines) and $l=1.6\times 10^5$ km (dashed-dot lines).  Similarly to the case with $l=0.8 \times 10^5$km (Section 3.2.4), the radiative fluxes of  the longer loops also show the metallicity dependence that lower $Z$ gives larger $L_{\rm X}$ and $L_{\rm UV}$.
Furthermore, Figure 9 indicates that longer loops emit moderately stronger $L_{\rm X}$ and $L_{\rm UV}$ because they have larger volumes. 
  
\section{Summary \& Discussion}  
We studied the heating of the coronal loops of low-mass stars with $Z=(0 - 1)Z_{\odot}$, particularly focusing on the coronae of metal-free stars. Comparing the time-averaged properties of the simulated coronal loops with the different metallicities, we found that lower-metal cases show higher temperature and density.  The density of the metal-free corona is $\approx 40$ times larger than that of the solar-metallicity case because of the lack of the coolants except for H and He. The denser structures in lower metal stars lead the volumetric heating rate to be higher, and the differential emission measure also takes larger values. As a result, the UV and X-ray luminosities from the metal-free corona are several times higher than those from the solar-metallicity corona even though the cooling efficiency is considerably lower. 
These results can support the past observations that detected the strong X-ray from the corona of some metal poor stars (Ottmann et al. 1997).

The time-dependent behavior of the simulated coronal loops is also quite different for different metallicities. In all the metallicity cases, the formation of the high-temperature corona occurs in a quasi-periodic manner; the steady corona is suddenly destructed by the enhanced cooling, which is followed by the rapid recovery of the corona by the continuous heating owing to Alfv\'{e}nic waves. The average duration of this quasi-periodicity is longer for lower metallicity, because the cooling time is longer. 

It is reported that such repeated behavior of the heating and the catastrophic cooling is also observed in solar coronal loops (Schrijver 2011; Kamio et al. 2011), which indicates that the quasi-periodic corona obtained in our simulations captures the basic nature of the coronal loops. However, we should note that the 1D treatment may slightly exaggerate the quasi-periodicity because propagating waves are confined in the flux tube without leaking out of it. A recent 3D simulation for the solar coronal loop (Matsumoto 2018) shows that the coronal temperature is kept with relatively small fluctuations of the temperature, $(0.7-1.2)\times 10^6$ K.  

The 1D treatment also affects the heating process of the wave dissipation.  In our simulations, the wave dissipation occurs through the nonlinear mode conversion that the incompressible Alfv\'{e}n waves turn into the compressible waves. However, in realistic situations that incorporate multi-dimensional effects, other dissipation channels of Alfv\'{e}n waves are also important. Cascading Alfv\'{e}nic turbulence (Goldreich \& Sridhar 1995), which has been highlighted in driving the solar wind (Mattaeus et al.1999; Cranmer et al. 2007; Shoda et al.2019), is supposed to operate in the heating of closed  loops (Matsumoto 2018). Phase mixing between Alfv\'{e}n waves that propagate neighboring field lines also contributes to the wave dissipation (Heyvaerts \& Priest 1983; De Groof \& Goossens 2002; Ofman \& Aschwanden 2002). 
When the driven frequency at the footpoints is comparable to the Alfv\'{e}n frequency of coronal loops, the resonant absorption of Alfv\'{e}nic waves is also proposed as a viable mechanism that heats up the coronae (Erd\'{e}lyi \& Goossens 1995; Erd\'{e}lyi 1998; Ofman et al. 1998). In realistic situations, these processes will additionally contribute to the dissipation of Alfv\'{e}n waves, while the nonlinear mode conversion to compressible waves is expected to be suppressed (Matsumoto \& Suzuki 2012; 2014). To solve the wave heating accurately, we need to incorporate such dissipation processes, which requires the multi-dimensional simulations and may modify our results.

We also performed the simulations of the longer loops with $l=(1.1,1.6)\times 10^5$ km than the standard case with $l=0.8\times 10^5$ km. The temperature in the corona is higher for longer loops because the density at the loop top located at a higher altitude is lower and the radiative cooling is less efficient. The UV and X-ray luminosities are also higher for longer loops because the emissions are from the larger volume. 

In addition to the 1D approximation, we  assumed the same set of the parameters, $T_{\rm eff}$, $\rho_{\rm ph}$,  $B_{\rm ph}$, and $\delta v$ at the footpoints of the simulated stars with the different metallicities so as to pick out the effect of the metallicity. An alternative way is to set the values at the realistic photosphere that takes into account the effect of different metallicity; the photosphere tends to be located at a deeper position in a lower-metal star because the opacity is smaller, and then, the photospheric density is higher (Suzuki 2018). Even if this prescription is adopted, there are still large uncertainties left in $B_{\rm ph}$. Therefore, it is important to study coronal properties in a wide range of the magnetic field strength.

Before closing the paper, we would like to mention a possible contribution from low-mass first stars to the cosmic reionization. EUV and probably soft X-ray radiations are regarded as reliable candidates of the reinization. There are several potential sources discussed in recent studies, such as active galactic nuclei (Shankar \& Mathur 2007; Volonteri \& Gnedin 2009; Glikman et al. 2011), X-ray binaries (Mirabel et al. 2011; Fialkov et al. 2017), mini-quasars (Madau et al. 2004; Fialkov et al. 2017) and dark matter annihilation (Belikov \& Hooper 2009), although their relative contributions are still under debate. The EUV and soft X-ray radiations from low-mass metal-free stars can be a potentially strong candidate if the sufficient number of such low-mass first stars were formed. In our future studies, we pursue the reionization in the high-$z$ universe by the EUV and soft X-ray radiations from low-mass first stars.\\

This work was supported by Grants-in-Aid for Scientific Research from the MEXT of Japan, 17H01105.

\small
\bibliographystyle{apj}
\nocite{*}
\bibliography{ref}

\begin{thebibliography}{}
\expandafter\ifx\csname natexlab\endcsname\relax\def\natexlab#1{#1}\fi

\bibitem[{{Alfv{\'e}n}(1947)}]{1947MNRAS.107..211A}
{Alfv{\'e}n}, H. 1947, \mnras, 107, 211

\bibitem[{{Anderson} \& {Athay}(1989)}]{1989ApJ...346.1010A}
{Anderson}, L.~S., \& {Athay}, R.~G. 1989, \apj, 346, 1010

\bibitem[{{Aoki} {et~al.}(2006){Aoki}, {Frebel}, {Christlieb}, {Norris},
  {Beers}, {Minezaki}, {Barklem}, {Honda}, {Takada-Hidai}, {Asplund}, {Ryan},
  {Tsangarides}, {Eriksson}, {Steinhauer}, {Deliyannis}, {Nomoto}, {Fujimoto},
  {Ando}, {Yoshii}, \& {Kajino}}]{2006ApJ...639..897A}
{Aoki}, W., {Frebel}, A., {Christlieb}, N., {et~al.} 2006, \apj, 639, 897

\bibitem[{{Athay}(1976)}]{1976ASSL...53.....A}
{Athay}, R.~G., ed. 1976, Astrophysics and Space Science Library, Vol.~53, {The
  solar chromosphere and corona: Quiet sun}

\bibitem[{{Belikov} \& {Hooper}(2009)}]{2009PhRvD..80c5007B}
{Belikov}, A.~V., \& {Hooper}, D. 2009, \prd, 80, 035007

\bibitem[{{Bromm} {et~al.}(2002){Bromm}, {Coppi}, \&
  {Larson}}]{2002ApJ...564...23B}
{Bromm}, V., {Coppi}, P.~S., \& {Larson}, R.~B. 2002, \apj, 564, 23

\bibitem[{{Chiaki} {et~al.}(2016){Chiaki}, {Yoshida}, \&
  {Hirano}}]{2016MNRAS.463.2781C}
{Chiaki}, G., {Yoshida}, N., \& {Hirano}, S. 2016, \mnras, 463, 2781

\bibitem[{{Christlieb} {et~al.}(2004){Christlieb}, {Gustafsson}, {Korn},
  {Barklem}, {Beers}, {Bessell}, {Karlsson}, \&
  {Mizuno-Wiedner}}]{2004ApJ...603..708C}
{Christlieb}, N., {Gustafsson}, B., {Korn}, A.~J., {et~al.} 2004, \apj, 603,
  708

\bibitem[{{Clark} {et~al.}(2011){Clark}, {Glover}, {Smith}, {Greif}, {Klessen},
  \& {Bromm}}]{2011Sci...331.1040C}
{Clark}, P.~C., {Glover}, S.~C.~O., {Smith}, R.~J., {et~al.} 2011, Science,
  331, 1040

\bibitem[{{Cranmer} {et~al.}(2007){Cranmer}, {van Ballegooijen}, \&
  {Edgar}}]{2007ApJS..171..520C}
{Cranmer}, S.~R., {van Ballegooijen}, A.~A., \& {Edgar}, R.~J. 2007, \apjs,
  171, 520

\bibitem[{{De Groof} \& {Goossens}(2002)}]{2002A&A...386..691D}
{De Groof}, A., \& {Goossens}, M. 2002, \aap, 386, 691

\bibitem[{{Edl{\'e}n}(1943)}]{1943ZA.....22...30E}
{Edl{\'e}n}, B. 1943, \zap, 22, 30

\bibitem[{{Erdelyi}(1998)}]{1998SoPh..180..213E}
{Erdelyi}, R. 1998, \solphys, 180, 213

\bibitem[{{Erdelyi} \& {Goossens}(1995)}]{1995A&A...294..575E}
{Erdelyi}, R., \& {Goossens}, M. 1995, \aap, 294, 575

\bibitem[{{Fialkov} {et~al.}(2017){Fialkov}, {Cohen}, {Barkana}, \&
  {Silk}}]{2017MNRAS.464.3498F}
{Fialkov}, A., {Cohen}, A., {Barkana}, R., \& {Silk}, J. 2017, \mnras, 464,
  3498

\bibitem[{{Glikman} {et~al.}(2011){Glikman}, {Djorgovski}, {Stern}, {Dey},
  {Jannuzi}, \& {Lee}}]{2011ApJ...728L..26G}
{Glikman}, E., {Djorgovski}, S.~G., {Stern}, D., {et~al.} 2011, \apjl, 728, L26

\bibitem[{{Goldreich} \& {Sridhar}(1995)}]{1995ApJ...438..763G}
{Goldreich}, P., \& {Sridhar}, S. 1995, \apj, 438, 763

\bibitem[{{Goldstein}(1978)}]{1978ApJ...219..700G}
{Goldstein}, M.~L. 1978, \apj, 219, 700

\bibitem[{{Greif} {et~al.}(2012){Greif}, {Bromm}, {Clark}, {Glover}, {Smith},
  {Klessen}, {Yoshida}, \& {Springel}}]{2012MNRAS.424..399G}
{Greif}, T.~H., {Bromm}, V., {Clark}, P.~C., {et~al.} 2012, \mnras, 424, 399

\bibitem[{{Grotrian}(1939)}]{1939NW.....27..555G}
{Grotrian}, W. 1939, Naturwissenschaften, 27, 555

\bibitem[{{G{\"u}del}(2007)}]{2007MmSAI..78..422G}
{G{\"u}del}, M. 2007, \memsai, 78, 422

\bibitem[{{Heyvaerts} \& {Priest}(1983)}]{1983A&A...117..220H}
{Heyvaerts}, J., \& {Priest}, E.~R. 1983, \aap, 117, 220

\bibitem[{{Hirano} {et~al.}(2014){Hirano}, {Hosokawa}, {Yoshida}, {Umeda},
  {Omukai}, {Chiaki}, \& {Yorke}}]{2014ApJ...781...60H}
{Hirano}, S., {Hosokawa}, T., {Yoshida}, N., {et~al.} 2014, \apj, 781, 60

\bibitem[{{Hollweg} {et~al.}(1982){Hollweg}, {Jackson}, \&
  {Galloway}}]{1982SoPh...75...35H}
{Hollweg}, J.~V., {Jackson}, S., \& {Galloway}, D. 1982, \solphys, 75, 35

\bibitem[{{Kamio} {et~al.}(2011){Kamio}, {Peter}, {Curdt}, \&
  {Solanki}}]{2011A&A...532A..96K}
{Kamio}, S., {Peter}, H., {Curdt}, W., \& {Solanki}, S.~K. 2011, \aap, 532, A96

\bibitem[{{Keller} {et~al.}(2014){Keller}, {Bessell}, {Frebel}, {Casey},
  {Asplund}, {Jacobson}, {Lind}, {Norris}, {Yong}, {Heger}, {Magic}, {da
  Costa}, {Schmidt}, \& {Tisserand}}]{2014Natur.506..463K}
{Keller}, S.~C., {Bessell}, M.~S., {Frebel}, A., {et~al.} 2014, \nat, 506, 463

\bibitem[{{Komiya} {et~al.}(2015){Komiya}, {Suda}, \&
  {Fujimoto}}]{2015ApJ...808L..47K}
{Komiya}, Y., {Suda}, T., \& {Fujimoto}, M.~Y. 2015, \apjl, 808, L47

\bibitem[{{Kuin} \& {Martens}(1982)}]{1982A&A...108L...1K}
{Kuin}, N.~P.~M., \& {Martens}, P.~C.~H. 1982, \aap, 108, L1

\bibitem[{{Machida} \& {Doi}(2013)}]{2013MNRAS.435.3283M}
{Machida}, M.~N., \& {Doi}, K. 2013, \mnras, 435, 3283

\bibitem[{{Madau} {et~al.}(2004){Madau}, {Rees}, {Volonteri}, {Haardt}, \&
  {Oh}}]{2004ApJ...604..484M}
{Madau}, P., {Rees}, M.~J., {Volonteri}, M., {Haardt}, F., \& {Oh}, S.~P. 2004,
  \apj, 604, 484

\bibitem[{{Matsumoto}(2018)}]{2018MNRAS.476.3328M}
{Matsumoto}, T. 2018, \mnras, 476, 3328

\bibitem[{{Matsumoto} \& {Suzuki}(2012)}]{2012ApJ...749....8M}
{Matsumoto}, T., \& {Suzuki}, T.~K. 2012, \apj, 749, 8

\bibitem[{{Matsumoto} \& {Suzuki}(2014)}]{2014MNRAS.440..971M}
---. 2014, \mnras, 440, 971

\bibitem[{Matthaeus {et~al.}(1999)Matthaeus, Zank, Oughton, Mullan, \&
  Dmitruk}]{Matthaeus_1999}
Matthaeus, W.~H., Zank, G.~P., Oughton, S., Mullan, D.~J., \& Dmitruk, P. 1999,
  The Astrophysical Journal, 523, L93

\bibitem[{{McIntosh} {et~al.}(2011){McIntosh}, {de Pontieu}, {Carlsson},
  {Hansteen}, {Boerner}, \& {Goossens}}]{2011Natur.475..477M}
{McIntosh}, S.~W., {de Pontieu}, B., {Carlsson}, M., {et~al.} 2011, \nat, 475,
  477

\bibitem[{{Mendoza-Brice{\~n}o} {et~al.}(2005){Mendoza-Brice{\~n}o},
  {Sigalotti}, \& {Erd{\'e}lyi}}]{2005ApJ...624.1080M}
{Mendoza-Brice{\~n}o}, C.~A., {Sigalotti}, L.~D.~G., \& {Erd{\'e}lyi}, R. 2005,
  \apj, 624, 1080

\bibitem[{{Mirabel} {et~al.}(2011){Mirabel}, {Dijkstra}, {Laurent}, {Loeb}, \&
  {Pritchard}}]{2011A&A...528A.149M}
{Mirabel}, I.~F., {Dijkstra}, M., {Laurent}, P., {Loeb}, A., \& {Pritchard},
  J.~R. 2011, \aap, 528, A149

\bibitem[{{Moriyasu} {et~al.}(2004){Moriyasu}, {Kudoh}, {Yokoyama}, \&
  {Shibata}}]{2004ApJ...601L.107M}
{Moriyasu}, S., {Kudoh}, T., {Yokoyama}, T., \& {Shibata}, K. 2004, \apjl, 601,
  L107

\bibitem[{{M{\"u}ller} {et~al.}(2003){M{\"u}ller}, {Hansteen}, \&
  {Peter}}]{2003A&A...411..605M}
{M{\"u}ller}, D.~A.~N., {Hansteen}, V.~H., \& {Peter}, H. 2003, \aap, 411, 605

\bibitem[{{M{\"u}ller} {et~al.}(2004){M{\"u}ller}, {Peter}, \&
  {Hansteen}}]{2004A&A...424..289M}
{M{\"u}ller}, D.~A.~N., {Peter}, H., \& {Hansteen}, V.~H. 2004, \aap, 424, 289

\bibitem[{{Ofman} \& {Aschwanden}(2002)}]{2002ApJ...576L.153O}
{Ofman}, L., \& {Aschwanden}, M.~J. 2002, \apjl, 576, L153

\bibitem[{{Ofman} {et~al.}(1998){Ofman}, {Klimchuk}, \&
  {Davila}}]{1998ApJ...493..474O}
{Ofman}, L., {Klimchuk}, J.~A., \& {Davila}, J.~M. 1998, \apj, 493, 474

\bibitem[{{Okamoto} {et~al.}(2007){Okamoto}, {Tsuneta}, {Berger}, {Ichimoto},
  {Katsukawa}, {Lites}, {Nagata}, {Shibata}, {Shimizu}, {Shine}, {Suematsu},
  {Tarbell}, \& {Title}}]{2007Sci...318.1577O}
{Okamoto}, T.~J., {Tsuneta}, S., {Berger}, T.~E., {et~al.} 2007, Science, 318,
  1577

\bibitem[{{Omukai} \& {Palla}(2001)}]{2001ApJ...561L..55O}
{Omukai}, K., \& {Palla}, F. 2001, \apjl, 561, L55

\bibitem[{{O'Shea} \& {Norman}(2007)}]{2007ApJ...654...66O}
{O'Shea}, B.~W., \& {Norman}, M.~L. 2007, \apj, 654, 66

\bibitem[{{Ottmann} {et~al.}(1997){Ottmann}, {Fleming}, \&
  {Pasquini}}]{1997A&A...322..785O}
{Ottmann}, R., {Fleming}, T.~A., \& {Pasquini}, L. 1997, \aap, 322, 785

\bibitem[{{Parker}(1958{\natexlab{a}})}]{1958PhFl....1..171P}
{Parker}, E.~N. 1958{\natexlab{a}}, Physics of Fluids, 1, 171

\bibitem[{{Parker}(1958{\natexlab{b}})}]{1958ApJ...128..677P}
---. 1958{\natexlab{b}}, \apj, 128, 677

\bibitem[{{Parker}(1988)}]{1988ApJ...330..474P}
---. 1988, \apj, 330, 474

\bibitem[{{Ribas} {et~al.}(2005){Ribas}, {Guinan}, {G{\"u}del}, \&
  {Audard}}]{2005ApJ...622..680R}
{Ribas}, I., {Guinan}, E.~F., {G{\"u}del}, M., \& {Audard}, M. 2005, \apj, 622,
  680

\bibitem[{{Richard} {et~al.}(2002){Richard}, {Michaud}, {Richer}, {Turcotte},
  {Turck-Chi{\`e}ze}, \& {VandenBerg}}]{2002ApJ...568..979R}
{Richard}, O., {Michaud}, G., {Richer}, J., {et~al.} 2002, \apj, 568, 979

\bibitem[{{Sano} \& {Miyama}(1999)}]{1999ApJ...515..776S}
{Sano}, T., \& {Miyama}, S.~M. 1999, \apj, 515, 776

\bibitem[{{Schrijver}(2001)}]{2001SoPh..198..325S}
{Schrijver}, C.~J. 2001, \solphys, 198, 325

\bibitem[{{Shankar} \& {Mathur}(2007)}]{2007ApJ...660.1051S}
{Shankar}, F., \& {Mathur}, S. 2007, \apj, 660, 1051

\bibitem[{{Shimizu}(1995)}]{1995PASJ...47..251S}
{Shimizu}, T. 1995, \pasj, 47, 251

\bibitem[{{Shoda} {et~al.}(2019){Shoda}, {Suzuki}, {Asgari-Targhi}, \&
  {Yokoyama}}]{2019ApJ...880L...2S}
{Shoda}, M., {Suzuki}, T.~K., {Asgari-Targhi}, M., \& {Yokoyama}, T. 2019,
  \apjl, 880, L2

\bibitem[{{Shoda} {et~al.}(2018){Shoda}, {Yokoyama}, \&
  {Suzuki}}]{2018ApJ...860...17S}
{Shoda}, M., {Yokoyama}, T., \& {Suzuki}, T.~K. 2018, \apj, 860, 17

\bibitem[{{Skumanich}(1972)}]{1972ApJ...171..565S}
{Skumanich}, A. 1972, \apj, 171, 565

\bibitem[{{Stone} \& {Norman}(1992)}]{1992ApJS...80..791S}
{Stone}, J.~M., \& {Norman}, M.~L. 1992, \apjs, 80, 791

\bibitem[{{Suematsu} {et~al.}(1995){Suematsu}, {Ichimoto}, \&
  {Sakurai}}]{1995itsa.conf..413S}
{Suematsu}, Y., {Ichimoto}, K., \& {Sakurai}, T. 1995, in Infrared tools for
  solar astrophysics: What's next?, ed. J.~R. {Kuhn} \& M.~J. {Penn}, 413

\bibitem[{{Susa} {et~al.}(2014){Susa}, {Hasegawa}, \&
  {Tominaga}}]{2014ApJ...792...32S}
{Susa}, H., {Hasegawa}, K., \& {Tominaga}, N. 2014, \apj, 792, 32

\bibitem[{{Sutherland} \& {Dopita}(1993)}]{1993ApJS...88..253S}
{Sutherland}, R.~S., \& {Dopita}, M.~A. 1993, \apjs, 88, 253

\bibitem[{{Suzuki}(2018)}]{2018PASJ...70...34S}
{Suzuki}, T.~K. 2018, \pasj, 70, 34

\bibitem[{{Suzuki} \& {Inutsuka}(2005)}]{2005ApJ...632L..49S}
{Suzuki}, T.~K., \& {Inutsuka}, S.-i. 2005, \apjl, 632, L49

\bibitem[{{Suzuki} \& {Inutsuka}(2006)}]{2006JGRA..111.6101S}
{Suzuki}, T.~K., \& {Inutsuka}, S.-I. 2006, Journal of Geophysical Research
  (Space Physics), 111, A06101

\bibitem[{{Tanaka} {et~al.}(2017){Tanaka}, {Chiaki}, {Tominaga}, \&
  {Susa}}]{2017ApJ...844..137T}
{Tanaka}, S.~J., {Chiaki}, G., {Tominaga}, N., \& {Susa}, H. 2017, \apj, 844,
  137

\bibitem[{{Tanikawa} {et~al.}(2018){Tanikawa}, {Suzuki}, \&
  {Doi}}]{2018PASJ...70...80T}
{Tanikawa}, A., {Suzuki}, T.~K., \& {Doi}, Y. 2018, \pasj, 70, 80

\bibitem[{{Terasawa} {et~al.}(1986){Terasawa}, {Hoshino}, {Sakai}, \&
  {Hada}}]{1986JGR....91.4171T}
{Terasawa}, T., {Hoshino}, M., {Sakai}, J.-I., \& {Hada}, T. 1986, \jgr, 91,
  4171

\bibitem[{{Verdini} {et~al.}(2012){Verdini}, {Grappin}, \&
  {Velli}}]{2012A&A...538A..70V}
{Verdini}, A., {Grappin}, R., \& {Velli}, M. 2012, \aap, 538, A70

\bibitem[{{Volonteri} \& {Gnedin}(2009)}]{2009ApJ...703.2113V}
{Volonteri}, M., \& {Gnedin}, N.~Y. 2009, \apj, 703, 2113

\bibitem[{{Wood} {et~al.}(2005){Wood}, {Redfield}, {Linsky}, {M{\"u}ller}, \&
  {Zank}}]{2005ESASP.560..309W}
{Wood}, B.~E., {Redfield}, S., {Linsky}, J.~L., {M{\"u}ller}, H.-R., \& {Zank},
  G.~P. 2005, in ESA Special Publication, Vol. 560, 13th Cambridge Workshop on
  Cool Stars, Stellar Systems and the Sun, ed. F.~{Favata}, G.~A.~J. {Hussain},
  \& B.~{Battrick}, 309

\bibitem[{{Yi} {et~al.}(2001){Yi}, {Demarque}, {Kim}, {Lee}, {Ree}, {Lejeune},
  \& {Barnes}}]{2001ApJS..136..417Y}
{Yi}, S., {Demarque}, P., {Kim}, Y.-C., {et~al.} 2001, \apjs, 136, 417

\bibitem[{{Yi} {et~al.}(2003){Yi}, {Kim}, \& {Demarque}}]{2003ApJS..144..259Y}
{Yi}, S.~K., {Kim}, Y.-C., \& {Demarque}, P. 2003, \apjs, 144, 259

\bibitem[{{Yoshida} {et~al.}(2006){Yoshida}, {Omukai}, {Hernquist}, \&
  {Abel}}]{2006ApJ...652....6Y}
{Yoshida}, N., {Omukai}, K., {Hernquist}, L., \& {Abel}, T. 2006, \apj, 652, 6

\bibitem[{{Yoshii}(1981)}]{1981A&A....97..280Y}
{Yoshii}, Y. 1981, \aap, 97, 280

\end{thebibliography}


\end{document}